\documentclass[11pt]{article}
\usepackage{amssymb,amsmath,amsfonts}
\usepackage{graphicx}
\usepackage{graphics}
\usepackage{eepic,epsfig}

\begin{document}

\title{Electromagnetic Casimir densities for a wedge with a coaxial
cylindrical shell }
\author{A. A. Saharian\thanks{%
E-mail: saharian@ictp.it} \\
\\
\textit{Department of Physics, Yerevan State University, 1 Alex Manoogian Street,}\\
\textit{375025 Yerevan, Armenia}}
\maketitle

\begin{abstract}
Vacuum expectation values of the field square and the
energy-momentum tensor for the electromagnetic field are
investigated for the geometry of a wedge with a coaxal cylindrical
boundary. All boundaries are assumed to be perfectly conducting
and both regions inside and outside the shell are considered. By
using the generalized Abel-Plana formula, the vacuum expectation
values are presented in the form of the sum of two terms. The
first one corresponds to the geometry of the wedge without the
cylindrical shell and the second term is induced by the presence
of the shell. The vacuum energy density induced by the shell is
negative for the interior region and is positive for the exterior
region. The asymptotic behavior of the vacuum expectation values
are investigated in various limiting cases. It is shown that the
vacuum forces acting on the wedge sides due to the presence of the
cylindrical boundary are always attractive.
\end{abstract}

\bigskip


\bigskip

\section{Introduction}

The Casimir effect is among the most interesting macroscopic manifestations
of vacuum fluctuations. It have important implications on all scales, from
cosmological to subnuclear, and has become in recent decades an increasingly
popular topic in quantum field theory. In addition to its fundamental
interest the Casimir effect also plays an important role in the fabrication
and operation of nano- and micro-scale mechanical systems. The imposition of
boundary conditions on a quantum field leads to the modification of the
spectrum for the zero-point fluctuations and results in the shift in the
vacuum expectation values for physical quantities such as the energy density
and stresses. In particular, the confinement of quantum fluctuations causes
forces that act on constraining boundaries. The particular features of the
resulting vacuum forces depend on the nature of the quantum field, the type
of spacetime manifold, the boundary geometries and the specific boundary
conditions imposed on the field. Since the original work by Casimir \cite%
{Casi48} many theoretical and experimental works have been done on this
problem (see, e.g., \cite{Most97,Bord01,Milt02,Nest05} and references
therein). Many different approaches have been used: mode summation method
with combination of the zeta function regularization technique, Green
function formalism, multiple scattering expansions, heat-kernel series, etc.
Advanced field-theoretical methods have been developed for Casimir
calculations during the past years \cite{Bord96,Grah02,Scha98}. However,
there are still difficulties in both interpretation and renormalization of
the Casimir effect. Straightforward computations of geometry dependencies
are conceptually complicated, since relevant information is subtly encoded
in the fluctuations spectrum \cite{Scha98}. Analytic solutions can usually
be found only for highly symmetric geometries including planar, spherically
and cylindrically symmetric boundaries. Recently the Casimir energy has been
evaluated exactly for several less symmetric configurations of experimental
interest. These include a sphere in front of a plane and a cylinder in front
of a plane \cite{Bulg06}.

A great deal of attention received the investigations of quantum effects for
cylindrical boundaries. In addition to traditional problems of quantum
electrodynamics under the presence of material boundaries, the Casimir
effect for cylindrical geometries can also be important to the flux tube
models of confinement \cite{Fish87,Barb90} and for determining the structure
of the vacuum state in interacting field theories \cite{Ambj83}. The
calculation of the vacuum energy of electromagnetic field with boundary
conditions defined on a cylinder turned out to be technically a more
involved problem than the analogous one for a sphere. First the Casimir
energy of an infinite perfectly conducting cylindrical shell has been
calculated in Ref. \cite{Dera81} by introducing ultraviolet cutoff and later
the corresponding result was derived by zeta function technique \cite%
{Milt99,Gosd98,Lamb99}. The local characteristics of the corresponding
electromagnetic vacuum such as energy density and vacuum stresses are
considered in \cite{Sah1cyl} for the interior and exterior regions of a
conducting cylindrical shell, and in \cite{Sah2cyl} for the region between
two coaxial shells (see also \cite{Sahrev}). The vacuum forces acting on the
boundaries in the geometry of two cylinders are also considered in Refs.
\cite{Mazz02}. The scalar Casimir densities for a single and two coaxial
cylindrical shells with Robin boundary conditions are investigated in Refs.
\cite{Rome01,Saha06}. Less symmetric configuration of two eccentric
perfectly conducting cylinders is considered in Ref. \cite{Mazz06}. Vacuum
energy for a perfectly conducting cylinder of elliptical section is
evaluated in Ref. \cite{Kits06} by the mode summation method, using the
ellipticity as a perturbation parameter. The Casimir forces acting on two
parallel plates inside a conducting cylindrical shell are investigated in
Ref. \cite{Mara07}.

Aside from their own theoretical and experimental interest, the exactly
solvable problems with this type of boundaries are useful for testing the
validity of various approximations used to deal with more complicated
geometries. From this point of view the wedge with a coaxial cylindrical
boundary is an interesting system, since the geometry is nontrivial and it
includes two dynamical parameters, radius of the cylindrical shell and
opening angle of the wedge. This geometry is also interesting from the point
of view of general analysis for surface divergences in the expectation
values of local physical observables for boundaries with discontinuities.
The nonsmoothness of the boundary generates additional contributions to the
heat kernel coefficients (see, for instance, the discussion in \cite%
{Apps98,Dowk00,Nest03} and references therein). The present paper is
concerned with local analysis of the vacuum of the electromagnetic field
constrained to satisfy perfectly conducting boundary conditions on boundary
surfaces of a wedge with a coaxial cylindrical boundary. Namely, we will
study the vacuum expectation values of the field squares and the
energy-momentum tensor for the electromagnetic field for both regions inside
and outside the cylindrical shell. In addition to describing the physical
structure of the quantum field at a given point, the energy-momentum tensor
acts as the source of gravity in the Einstein equations. It therefore plays
an important role in modelling a self-consistent dynamics involving the
gravitational field. The vacuum expectation value of the square of the
electric field determines the electromagnetic force on a neutral polarizable
particle. Some most relevant investigations to the present paper are
contained in Refs. \cite{Most97,jphy,Deutsch,brevikI,Brev98,brevikII}, where
the geometry of a wedge without a cylindrical boundary is considered for a
conformally coupled scalar and electromagnetic fields in a four dimensional
spacetime. The total Casimir energy of a semi-circular infinite cylindrical
shell with perfectly conducting walls is considered in \cite{Nest01} by
using the zeta function technique. For a scalar field with an arbitrary
curvature coupling parameter the Wightman function, the vacuum expectation
values of the field square and the energy-momentum tensor in the geometry of
a wedge with an arbitrary opening angle and with a cylindrical boundary are
evaluated in \cite{Reza02,Saha05}. Note that, unlike the case of conformally
coupled fields, for a general coupling the vacuum energy-momentum tensor is
angle-dependent and diverges on the wedge sides. Our method here employs the
mode summation and is based on a variant of the generalized Abel-Plana
formula \cite{Saha87} (see also Refs. \cite{Sahrev,Saha06AP}). This enables
us to extract from the vacuum expectation values the parts due to a wedge
without the cylindrical shell and to present the parts induced by the shell
in terms of strongly convergent integrals. Note that the closely related
problem of the vacuum densities induced by a cylindrical boundary in the
geometry of a cosmic string is investigated in Refs. \cite{Beze06,Beze07}
for both scalar and electromagnetic fields.

We have organized the paper as follows. In the next section we
describe the structure of the modes for a wedge with a cylindrical
shell in the region inside the shell. By applying to the
corresponding mode sums the generalized Abel-Plana formula, we
evaluate the vacuum expectation values of the electric and
magnetic field square. Various limiting cases of the general
formulae are discussed. Section \ref{sec:EMTint} is devoted to the
investigation of the vacuum expectation values for the
energy-momentum tensor of the electromagnetic field in the region
inside the shell. The additional vacuum forces acting on the wedge
sides due to the presence of the cylindrical boundary are
evaluated. In section \ref{sec:exter} we consider the vacuum
densities for a wedge with the cylindrical shell in the exterior
region with respect to the shell. Formulae for the shell
contributions are derived and the corresponding surface
divergences are investigated. The vacuum forces acting on the
wedge sides are discussed. The main results are summarized and
discussed in section \ref{sec:Conc}.

\section{Vacuum expectation values of the field square inside a cylindrical
shell}

\label{sec:Inter}

\begin{figure}[tbph]
\begin{center}
\epsfig{figure=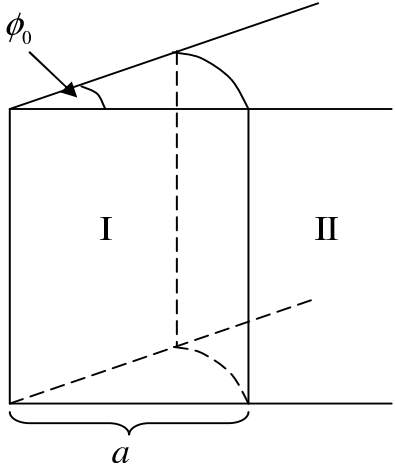,width=5.5cm,height=6.5cm}
\end{center}
\caption{Geometry of a wedge with a coaxial cylindrical boundary with radius
$a$.}
\label{fig1}
\end{figure}

Consider a wedge with the opening angle $\phi _{0}$ and with a coaxial
cylindrical boundary of radius $a$ (see figure \ref{fig1}) assuming that all
boundaries are perfectly conducting. In accordance with the problem
symmetry, in the discussion below the cylindrical coordinates $(r,\phi ,z)$
will be used. We are interested in the vacuum expectation values (VEVs) of
the field square and the energy-momentum tensor for the electromagnetic
field.
Expanding the field operator in terms of the creation and annihilation
operators and using the commutation relations, the VEV for a quantity $%
F\left\{ A_{i},A_{k}\right\} $ bilinear in the field can be presented in the
form of the mode-sum%
\begin{equation}
\langle 0|F\left\{ A_{i},A_{k}\right\} |0\rangle =\sum_{\alpha }F\left\{
A_{\alpha i},A_{\alpha k}^{\ast }\right\} ,  \label{VEVbil}
\end{equation}%
where $\left\{ A_{\alpha i},A_{\alpha k}^{\ast }\right\} $ is a complete set
of solutions of the classical field equations satisfying the boundary
conditions on the bounding surfaces and specified by a set of quantum
numbers $\alpha $.

In accordance with formula (\ref{VEVbil}), for the evaluation of the VEVs
for the square of the electric and magnetic fields and the energy-momentum
tensor, the corresponding eigenfunctions are needed. In this section we
consider the region inside the cylindrical shell (region I in figure \ref%
{fig1}). For the geometry under consideration there are two different types
of the eigenfunctions corresponding to the transverse magnetic (TM) and
transverse electric (TE) waves. In the discussion below we will specify
these modes by the index $\lambda =0$ and $\lambda =1$ for the TM and TE
waves respectively. In the Coulomb gauge, the vector potentials for the TM
and TE modes are given by the formulae%
\begin{equation}
\mathbf{A}_{\alpha }=\beta _{\alpha }\left\{
\begin{array}{cc}
(1/i\omega )\left( \gamma ^{2}\mathbf{e}_{3}+ik\nabla _{t}\right)
J_{q|m|}(\gamma r)\sin (qm\phi )\exp \left[ i\left( kz-\omega t\right) %
\right] , & \lambda =0 \\
-\mathbf{e}_{3}\times \nabla _{t}\left\{ J_{q|m|}(\gamma r)\cos (qm\phi
)\exp \left[ i\left( kz-\omega t\right) \right] \right\} , & \lambda =1%
\end{array}%
\right. ,  \label{Aalpha}
\end{equation}%
where $\mathbf{e}_{3}$ is the unit vector along the axis of the wedge, $%
\nabla _{t}$ is the part of the nabla operator transverse to this axis, $%
J_{\nu }(x)$ is the Bessel function of the first kind, and%
\begin{equation}
\omega ^{2}=\gamma ^{2}+k^{2},\;q=\pi /\phi _{0}.  \label{omega}
\end{equation}%
In Eq. (\ref{Aalpha}), $m=1,2,\ldots $ for $\lambda =0$ and $m=0,1,2,\ldots $
for $\lambda =1$. The normalization coefficient $\beta _{\alpha }$ is found
from the orthonormalization condition for the vector potential:%
\begin{equation}
\int dV\,\mathbf{A}_{\alpha }\cdot \mathbf{A}_{\alpha ^{\prime }}^{\ast }=%
\frac{2\pi }{\omega }\delta _{\alpha \alpha ^{\prime }},  \label{Anorm}
\end{equation}%
where the integration goes over the region inside the shell. From this
condition, by using the standard integral involving the square of the Bessel
function, one finds%
\begin{equation}
\beta _{\alpha }^{2}=\frac{4qT_{qm}(\gamma a)}{\pi \omega a\gamma }\delta
_{m},\;\delta _{m}=\left\{
\begin{array}{cc}
1/2, & m=0 \\
1, & m\neq 0%
\end{array}%
\right. ,  \label{betalf}
\end{equation}%
where we have introduced the notation
\begin{equation}
T_{\nu }(x)=x\left[ J_{\nu }^{^{\prime }2}(x)+(1-\nu ^{2}/x^{2})J_{\nu
}^{2}(x)\right] ^{-1}.  \label{Tnux}
\end{equation}

Eigenfunctions (\ref{Aalpha}) satisfy the standard boundary conditions for
the electric and magnetic fields, $\mathbf{n}\times \mathbf{E}=0$ and $%
\mathbf{n}\cdot \mathbf{B}=0$, on the wedge sides corresponding to $\phi =0$
and $\phi =\phi _{0}$, with $\mathbf{n}$\ being the normal to the boundary.
The eigenvalues for the quantum number $\gamma $ are determined by the
boundary conditions on the cylindrical shell. From the latter it follows
that these eigenvalues are solutions of the equation
\begin{equation}
J_{qm}^{(\lambda )}(\gamma a)=0,\quad \lambda =0,1,  \label{modes1}
\end{equation}%
where we use the notations $J_{\nu }^{(0)}(x)=J_{\nu }(x)$ and $J_{\nu
}^{(1)}(x)=J_{\nu }^{\prime }(x)$. We will denote the corresponding
eigenmodes by $\gamma a=j_{m,n}^{(\lambda )}$, $n=1,2,\ldots $, assuming
that the zeros $j_{m,n}^{(\lambda )}$ are arranged in ascending order.
Consequently, the eigenfunctions are specified by the set of quantum numbers
$\alpha =(k,m,\lambda ,n)$.

First we consider the VEVs of the squares of the electric and magnetic
fields inside the shell. Substituting the eigenfunctions (\ref{Aalpha}) into
the corresponding mode-sum formula, we find%
\begin{eqnarray}
\langle 0|F^{2}|0\rangle  &=&\frac{4q}{\pi a^{3}}\sideset{}{'}{\sum}%
_{m=0}^{\infty }\int_{-\infty }^{+\infty }dk\sum_{\lambda
=0,1}\sum_{n=1}^{\infty }\frac{j_{m,n}^{(\lambda )3}T_{qm}(j_{m,n}^{(\lambda
)})}{\sqrt{j_{m,n}^{(\lambda )2}+k^{2}a^{2}}}  \notag \\
&&\times g^{(\eta _{F\lambda })}[\Phi _{qm}^{(\lambda )}(\phi
),J_{qm}(j_{m,n}^{(\lambda )}r/a)],  \label{F2}
\end{eqnarray}%
where $F=E,B$ with $\eta _{E\lambda }=\lambda $, $\eta _{B\lambda
}=1-\lambda $, and the prime in the summation over $m$ means that the term $%
m=0$ should be halved. In formula (\ref{F2}), for a given function $f(x)$,
we have introduced the notations%
\begin{eqnarray}
g^{(0)}[\Phi (\phi ),f(x)] &=&(k^{2}r^{2}/x^{2})\left[ \Phi ^{2}(\phi
)f^{\prime 2}(x)+\Phi ^{\prime 2}(\phi )f^{2}(x)/x^{2}\right] +\Phi
^{2}(\phi )f^{2}(x),  \label{gnulam1} \\
g^{(1)}[\Phi (\phi ),f(x)] &=&(1+k^{2}r^{2}/x^{2})\left[ \Phi ^{2}(\phi
)f^{\prime 2}(x)+\Phi ^{\prime 2}(\phi )f^{2}(x)/x^{2}\right] ,
\label{gnulam2}
\end{eqnarray}%
and%
\begin{equation}
\Phi _{\nu }^{(\lambda )}(\phi )=\left\{
\begin{array}{cc}
\sin (\nu \phi ), & \lambda =0 \\
\cos (\nu \phi ), & \lambda =1%
\end{array}%
\right. .  \label{Philam}
\end{equation}%
The expressions (\ref{F2}) corresponding to the electric and magnetic fields
are divergent. They may be regularized introducing a cutoff function $\psi
_{\mu }(\omega )$ with the cutting parameter $\mu $ which makes the
divergent expressions finite and satisfies the condition $\psi _{\mu
}(\omega )\rightarrow 1$ for $\mu \rightarrow 0$. After the renormalization
the cutoff function is removed by taking the limit $\mu \rightarrow 0$. An
alternative way is to consider the product of the fields at different
spacetime points and to take the coincidence limit after the subtraction of
the corresponding Minkowskian part. Our approach here follows the first
method.

As we do not know the explicit expressions for the zeros $j_{m,n}^{(\lambda
)}$ as functions on $m$ and $n$, and the summand in formula (\ref{F2}) is
strongly oscillating function for large values of $m$ and $n$, this formula
is not convenient for the further evaluation of the VEVs of the field
square. In order to obtain an alternative representation, we apply to the
series over $n$ the generalized Abel-Plana summation formula \cite{Saha87}
(see also \cite{Saha06AP})
\begin{eqnarray}
\sum_{n=1}^{\infty }T_{qm}(j_{m,n}^{(\lambda )})f(j_{m,n}^{(\lambda )}) &=&%
\frac{1}{2}\int_{0}^{\infty }dx\,f(x)+\frac{\pi }{4}\underset{z=0}{\mathrm{%
Res}}f(z)\frac{Y_{qm}^{(\lambda )}(z)}{J_{qm}^{(\lambda )}(z)}  \notag \\
&&-\frac{1}{2\pi }\int_{0}^{\infty }dx\,\frac{K_{qm}^{(\lambda )}(x)}{%
I_{qm}^{(\lambda )}(x)}\left[ e^{-qm\pi i}f(e^{\pi i/2}x)+e^{qm\pi
i}f(e^{-\pi i/2}x)\right] ,  \label{sumformula}
\end{eqnarray}%
where $Y_{\nu }(z)$ is the Neumann function and $I_{\nu }(z)$, $K_{\nu }(z)$
are the modified Bessel functions. As it can be seen, for points away from
the shell the contribution to the VEVs coming from the second integral term
on the right-hand side of (\ref{sumformula}) is finite in the limit $\mu
\rightarrow 0$ and, hence, the cutoff function in this term can be safely
removed. As a result the VEVs can be written in the form%
\begin{equation}
\langle 0|F^{2}|0\rangle =\langle 0_{\mathrm{w}}|F^{2}|0_{\mathrm{w}}\rangle
+\left\langle F^{2}\right\rangle _{\mathrm{cyl}},  \label{F21}
\end{equation}%
where%
\begin{eqnarray}
\langle 0_{\mathrm{w}}|F^{2}|0_{\mathrm{w}}\rangle &=&\frac{q}{\pi }%
\sideset{}{'}{\sum}_{m=0}^{\infty }\int_{-\infty }^{+\infty
}dk\int_{0}^{\infty }d\gamma \,\frac{\gamma ^{3}\psi _{\mu }(\omega )}{\sqrt{%
\gamma ^{2}+k^{2}}}  \notag \\
&&\left\{ \left( 1+\frac{2k^{2}}{\gamma ^{2}}\right) \left[ J_{qm}^{\prime
2}(\gamma r)+\frac{q^{2}m^{2}}{\gamma ^{2}r^{2}}J_{qm}^{2}(\gamma r)\right]
+J_{qm}^{2}(\gamma r)\right.  \notag \\
&&\left. -(-1)^{\eta _{F1}}\cos (2qm\phi )\left[ J_{qm}^{\prime 2}(\gamma
r)-\left( 1+\frac{q^{2}m^{2}}{\gamma ^{2}r^{2}}\right) J_{qm}^{2}(\gamma r)%
\right] \right\} ,  \label{F2s}
\end{eqnarray}%
and%
\begin{equation}
\langle F^{2}\rangle _{\mathrm{cyl}}=\frac{8q}{\pi ^{2}}\sideset{}{'}{\sum}%
_{m=0}^{\infty }\int_{0}^{\infty }dk\sum_{\lambda =0,1}\int_{k}^{\infty
}dx\,x^{3}\,\frac{K_{qm}^{(\lambda )}(xa)}{I_{qm}^{(\lambda )}(xa)}\frac{%
G^{(\eta _{F\lambda })}[k,\Phi _{qm}^{(\lambda )}(\phi ),I_{qm}(xr)]}{\sqrt{%
x^{2}-k^{2}}}.  \label{F2b0}
\end{equation}%
In formula Eq. (\ref{F2b0}) we have introduced the notations%
\begin{eqnarray}
G^{(0)}[k,\Phi (\phi ),f(x)] &=&(k^{2}r^{2}/x^{2})\left[ \Phi ^{2}(\phi
)f^{\prime 2}(x)+\Phi ^{\prime 2}(\phi )f^{2}(x)/x^{2}\right] +\Phi
^{2}(\phi )f^{2}(x),  \label{Gnuj0} \\
G^{(1)}[k,\Phi (\phi ),f(x)] &=&(k^{2}r^{2}/x^{2}-1)\left[ \Phi ^{2}(\phi
)f^{\prime 2}(x)+\Phi ^{\prime 2}(\phi )f^{2}(x)/x^{2}\right] .
\label{Gnuj1}
\end{eqnarray}%
The second term on the right-hand side of Eq. (\ref{F21}) vanishes in the
limit $a\rightarrow \infty $ and the first one does not depend on $a$. Thus,
we can conclude that the term $\langle 0_{\mathrm{w}}|F^{2}|0_{\mathrm{w}%
}\rangle $ corresponds to the part in the VEVs\ when the cylindrical shell
is absent with the corresponding vacuum state $|0_{\mathrm{w}}\rangle $.
Hence, the application of the generalized Abel-Plana formula enables us to
extract from the VEVs the parts induced by the cylindrical shell without
specifying the cutoff function. In addition, these parts are presented in
terms of the exponentially convergent integrals.

First, let us concentrate on the part corresponding to the wedge without a
cylindrical shell. First of all we note that in Eq. (\ref{F2s}) the part
which does not depend on the angular coordinate $\phi $ is the same as in
the corresponding problem of the cosmic string geometry with the angle
deficit $2\pi -\phi _{0}$ (see Ref. \cite{Beze07}), which we will denote by $%
\langle 0_{\mathrm{s}}|F^{2}|0_{\mathrm{s}}\rangle $. For this part we have
\begin{eqnarray}
\langle 0_{\mathrm{s}}|F^{2}|0_{\mathrm{s}}\rangle &=&\frac{q}{\pi }%
\sideset{}{'}{\sum}_{m=0}^{\infty }\int_{-\infty }^{+\infty
}dk\int_{0}^{\infty }d\gamma \,\frac{\gamma ^{3}\psi _{\mu }(\omega )}{\sqrt{%
\gamma ^{2}+k^{2}}}  \notag \\
&&\times \left\{ \left( 1+\frac{2k^{2}}{\gamma ^{2}}\right) \left[
J_{qm}^{\prime 2}(\gamma r)+\frac{q^{2}m^{2}}{\gamma ^{2}r^{2}}%
J_{qm}^{2}(\gamma r)\right] +J_{qm}^{2}(\gamma r)\right\}  \notag \\
&=&\langle 0_{\mathrm{M}}|F^{2}|0_{\mathrm{M}}\rangle -\frac{%
(q^{2}-1)(q^{2}+11)}{180\pi r^{4}},  \label{F2sn}
\end{eqnarray}%
where $\langle 0_{\mathrm{M}}|F^{2}|0_{\mathrm{M}}\rangle $ is the part
corresponding to the Minkowskian spacetime without boundaries and in the
last expression we have removed the cutoff. To evaluate the part in (\ref%
{F2s}) which depends on $\phi $, we firstly consider the case when the
parameter $q$ is an integer. In this case the summation over $m$ can be done
by using the formula \cite{Prud86,Davi88}
\begin{equation}
\sideset{}{'}{\sum}_{m=0}^{\infty }\cos (2qm\phi )J_{qm}^{2}(y)=\frac{1}{2q}%
\sum_{l=0}^{q-1}J_{0}(2y\sin (\phi +\phi _{0}l)).  \label{rel1}
\end{equation}%
The formulae for the other series entering in Eq. (\ref{F2s}) are obtained
from (\ref{rel1}) taking the derivatives with respect to $\phi $ and $y$. In
particular, for the combination appearing in the angle-dependent part we
obtain%
\begin{equation}
\sideset{}{'}{\sum}_{m=0}^{\infty }\cos (2qm\phi )\left[ J_{qm}^{\prime
2}(y)-\left( 1+\frac{q^{2}m^{2}}{y^{2}}\right) J_{qm}^{2}(y)\right] =-\frac{1%
}{q}\sum_{l=0}^{q-1}J_{1}^{\prime }(2y\sin (\phi +\phi _{0}l)).
\label{rel1n}
\end{equation}%
Substituting this in formula (\ref{F2s}), the integrals remained are
evaluated by introducing polar coordinates in the $(k,\gamma )$-plane. In
this way one finds%
\begin{equation}
\langle 0_{\mathrm{w}}|F^{2}|0_{\mathrm{w}}\rangle =\langle 0_{\mathrm{s}%
}|F^{2}|0_{\mathrm{s}}\rangle -\frac{3(-1)^{\eta _{F1}}}{4\pi r^{4}}%
\sum_{l=0}^{q-1}\sin ^{-4}(\phi +l\pi /q).  \label{F2w}
\end{equation}%
The sum on the right hand-side of this formula is evaluated by the double
differentiation of the relation \cite{Prud86}%
\begin{equation}
\sum_{l=0}^{q-1}\cos ^{-2}(x+l\pi /q)=q^{2}\sin ^{-2}(qx+q\pi /2).
\label{relsum}
\end{equation}%
Finally, for the renormalised VEVs\ of the field square in the geometry of a
wedge without a cylindrical boundary we find%
\begin{equation}
\langle F^{2}\rangle _{\mathrm{w},\mathrm{ren}}=-\frac{(q^{2}-1)(q^{2}+11)}{%
180\pi r^{4}}-\frac{(-1)^{\eta _{F1}}q^{2}}{2\pi r^{4}\sin ^{2}(q\phi )}%
\left[ \frac{3q^{2}}{2\sin ^{2}(q\phi )}+1-q^{2}\right] ,  \label{F2wren}
\end{equation}%
with $\eta _{E1}=1$ and $\eta _{B1}=0$. Though we have derived this formula
for integer values of the parameter $q$, by the analytic continuation it is
valid for non-integer values of this parameter as well. The expression on
the right of formula (\ref{F2wren}) is invariant under the replacement $\phi
\rightarrow \phi _{0}-\phi $ and, as we could expect, the VEVs are symmetric
with respect to the half-plane $\phi =\phi _{0}/2$. Formula (\ref{F2wren})
for $F=E$ was derived in Ref. \cite{Brev98} within the framework of
Schwinger's source theory. For $q=1$ from formula (\ref{F2wren}) as a
special case we obtain the renormalized VEVs of the field square for a
conducting plate. In this case $x=r\sin \phi $ is the distance from the
plate and one has%
\begin{equation}
\langle F^{2}\rangle _{\mathrm{pl},\mathrm{ren}}=-\frac{3(-1)^{\eta _{F1}}}{%
4\pi x^{4}}.  \label{F2plate}
\end{equation}%
Another special case $q=1/2$ corresponds to the geometry of a half-plane. In
(\ref{F2wren}) taking the limit $r\rightarrow \infty $, with $x_{0}=r\phi
_{0}$ being fixed, we obtain the corresponding results in the region between
two parallel plates located at the points $x=0$ and $x=x_{0}$:%
\begin{equation}
\langle F^{2}\rangle _{2\mathrm{pl},\mathrm{ren}}=-\frac{\pi ^{3}}{%
180x_{0}^{4}}-\frac{(-1)^{\eta _{F1}}\pi ^{3}}{2x_{0}^{4}\sin ^{2}(\pi
x/x_{0})}\left[ \frac{3}{2\sin ^{2}(\pi x/x_{0})}-1\right] .
\label{F22plate}
\end{equation}

Now, we turn to the investigation of the parts in the VEVs of the field
square induced by the cylindrical boundary and given by formula (\ref{F2b0}%
). By using the formula%
\begin{equation}
\int_{0}^{\infty }dk\,k^{m}\int_{k}^{\infty }dx\,\frac{xf(x)}{\sqrt{%
x^{2}-k^{2}}}=\frac{\sqrt{\pi }\Gamma \left( \frac{m+1}{2}\right) }{2\Gamma
\left( \frac{m}{2}+1\right) }\int_{0}^{\infty }dx\,x^{m+1}f(x),
\label{Hash1}
\end{equation}%
these parts are presented in the form
\begin{equation}
\langle F^{2}\rangle _{\mathrm{cyl}}=\frac{2q}{\pi }\sideset{}{'}{\sum}%
_{m=0}^{\infty }\sum_{\lambda =0,1}\int_{0}^{\infty }dx\,x^{3}\frac{%
K_{qm}^{(\lambda )}(xa)}{I_{qm}^{(\lambda )}(xa)}G^{(\eta _{F\lambda
})}[\Phi _{qm}^{(\lambda )}(\phi ),I_{qm}(xr)].  \label{F2cyl}
\end{equation}%
Here, for given functions $f(x)$ and $\Phi (\phi )$,\ we have introduced the
notations%
\begin{eqnarray}
G^{(0)}[\Phi (\phi ),f(x)] &=&\Phi ^{2}(\phi )f^{\prime 2}(x)+\Phi ^{\prime
2}(\phi )f^{2}(x)/x^{2}+2\Phi ^{2}(\phi )f^{2}(x),  \label{Gnujtilde0} \\
G^{(1)}[\Phi (\phi ),f(x)] &=&-\Phi ^{2}(\phi )f^{\prime 2}(x)-\Phi ^{\prime
2}(\phi )f^{2}(x)/x^{2}.  \label{Gnujtilde1}
\end{eqnarray}%
As we see the parts in the VEVs induced by the cylindrical shell are
symmetric with respect to the half-plane $\phi =\phi _{0}/2$.

The expression in the right-hand side of (\ref{F2cyl}) is finite for $0<r<a$
including the points on the wedge sides, and diverges on the shell. To find
the leading term in the corresponding asymptotic expansion, we note that
near the shell the main contribution comes from large values of $m$. By
using the uniform asymptotic expansions of the modified Bessel functions
(see, for instance, \cite{hand}) for large values of the order, up to the
leading order, for the points $a-r\ll a|\sin \phi |,a|\sin (\phi _{0}-\phi
)| $ we find%
\begin{equation}
\langle F^{2}\rangle _{\mathrm{cyl}}\approx -\frac{3(-1)^{\eta _{F1}}}{4\pi
(a-r)^{4}}.  \label{E2binnear}
\end{equation}%
For the points near the edges $(r=a,\phi =0,\phi _{0})$ the leading terms in
the corresponding asymptotic expansions are the same as for the geometry of
a wedge with the opening angle $\phi _{0}=\pi /2$. The leading terms given
by formula (\ref{E2binnear}) are the same as for the geometry of a single
plate (see (\ref{F2plate})). They do not depend on $\phi _{0}$ and have
opposite signs for the electric and magnetic fields. In particular, the
leading terms are cancelled in the evaluation of the vacuum energy density.
Surface divergences originate in the unphysical nature of perfect conductor
boundary conditions and are well-known in quantum field theory with
boundaries. In reality the expectation values will attain a limiting value
on the conductor surface, which will depend on the molecular details of the
conductor. From the formulae given above it follows that the main
contribution to $\langle F^{2}\rangle _{\mathrm{cyl}}$ are due to the
frequencies $\omega \lesssim (a-r)^{-1}$. Hence, we expect that formula (\ref%
{F2cyl}) is valid for real conductors up to distances $r$ for which $%
(a-r)^{-1}\ll \omega _{0}$, with $\omega _{0}$ being the characteristic
frequency, such that for $\omega >\omega _{0}$ the conditions for perfect
conductivity fail.

Near the edge $r=0$, assuming that $r/a\ll 1$, the asymptotic behavior of
the part induced in the VEVs of the field square by the cylindrical shell
depends on the parameter $q$. For $q>1+\eta _{F1}$, the dominant
contribution comes from the lowest mode $m=0$ and to the leading order one
has%
\begin{equation}
\langle F^{2}\rangle _{\mathrm{cyl}}\approx -(-1)^{\eta _{F1}}\frac{%
2^{1-\eta _{F1}}q}{\pi a^{4}}\left( \frac{r}{2a}\right) ^{2\eta
_{F1}}\int_{0}^{\infty }dx\,x^{3}\frac{K_{1}(x)}{I_{1}(x)}.
\label{F2nearcent1}
\end{equation}%
In this case the quantity $\langle B^{2}\rangle _{\mathrm{cyl}}$ takes a
finite limiting value on the edge $r=0$, whereas $\langle E^{2}\rangle _{%
\mathrm{cyl}}$ vanishes as $r^{2}$. For $q<1+\eta _{F1}$ the main
contribution comes form the mode with $m=1$ and the shell-induced parts
diverge on the edge $r=0$. The leading terms are given by the formula%
\begin{equation}
\langle F^{2}\rangle _{\mathrm{cyl}}\approx -\frac{(-1)^{\eta
_{F1}}q(r/a)^{2(q-1)}}{2^{2q-1}\pi \Gamma ^{2}(q)a^{4}}\int_{0}^{\infty
}dx\,x^{2q+1}\left[ \frac{K_{q}(x)}{I_{q}(x)}-\frac{K_{q}^{\prime }(x)}{%
I_{q}^{\prime }(x)}\right] .  \label{F2nearcent2}
\end{equation}%
As for the points near the shell, here the leading divergences in the VEVs
of the electric and magnetic fields are cancelled in the evaluation of the
vacuum energy density. For $q=1+\eta _{F1}$ the main contribution comes from
the modes $m=0,1$ and the corresponding asymptotic behavior is obtained by
summing the right-hand sides of Eqs. (\ref{F2nearcent1}) and (\ref%
{F2nearcent2}). In accordance with (\ref{F2wren}), near the edge $r=0$ the
total VEV is dominated by the part coming from the wedge without the
cylindrical shell. Here we have considered the VEVs for the field square.
The VEVs for the bilinear products of the fields at different spacetime
points may be evaluated in a similar way.

Now, we turn to the investigation of the behavior of the VEVs induced by the
cylindrical boundary in the limit $q\gg 1$. In this limit the order of the
modified Bessel functions is large for $m\neq 0$. By using the corresponding
asymptotic formulae it can be seen that the contribution of these terms is
suppressed by the factor $\exp [-2qm\ln (a/r)]$. As a result, the main
contribution comes from the lowest mode $m=0$ and the VEVs induced by the
cylindrical shell are proportional to $q$. Note that in this limit the part
corresponding to the wedge without the cylindrical shell behaves as $q^{4}$.

\section{Vacuum energy-momentum tensor inside the cylindrical shell}

\label{sec:EMTint}

Now let us consider the VEV of the energy-momentum tensor in the region
inside the cylindrical shell. Substituting the eigenfunctions (\ref{Aalpha})
into the corresponding mode-sum formula, for the non-zero components we
obtain (no summation over $i$)%
\begin{eqnarray}
\langle 0|T_{i}^{i}|0\rangle &=&\frac{q}{2\pi ^{2}a^{3}}\sideset{}{'}{\sum}%
_{m=0}^{\infty }\int_{-\infty }^{+\infty }dk\sum_{\lambda
=0,1}\sum_{n=1}^{\infty }\frac{j_{m,n}^{(\lambda )3}T_{qm}(j_{m,n}^{(\lambda
)})}{\sqrt{j_{m,n}^{(\lambda )2}+k^{2}a^{2}}}f^{(i)}[\Phi _{qm}^{(\lambda
)}(\phi ),J_{qm}(j_{m,n}^{(\lambda )}r/a)],  \label{Tik} \\
\langle 0|T_{2}^{1}|0\rangle &=&-\frac{q^{2}}{4\pi ^{2}a}\frac{\partial }{%
\partial r}\sideset{}{'}{\sum}_{m=0}^{\infty }m\sin (2qm\phi )\int_{-\infty
}^{+\infty }dk\sum_{\lambda =0,1}(-1)^{\lambda }  \notag \\
&&\times \sum_{n=1}^{\infty }\frac{j_{m,n}^{(\lambda
)}T_{qm}(j_{m,n}^{(\lambda )})}{\sqrt{j_{m,n}^{(\lambda )2}+k^{2}a^{2}}}%
J_{qm}^{2}(j_{m,n}^{(\lambda )}r/a),  \label{T21}
\end{eqnarray}%
where we have introduced the notations%
\begin{eqnarray}
f^{(j)}[\Phi (\phi ),f(x)] &=&(-1)^{i}\left( 2k^{2}/\gamma ^{2}+1\right)
\left[ \Phi ^{2}(\phi )f^{\prime 2}(x)+\Phi ^{\prime 2}(\phi )f^{2}(x)/y^{2}%
\right] +\Phi ^{2}(\phi )f^{2}(x),  \label{f0} \\
f^{(l)}[\Phi (\phi ),f(x)] &=&(-1)^{l}\Phi ^{2}(\phi )f^{\prime 2}(x)-\left[
\Phi ^{2}(\phi )+(-1)^{l}\Phi ^{\prime 2}(\phi )/x^{2}\right] f^{2}(x),
\label{fi}
\end{eqnarray}%
with $j=0,3$ and $l=1,2$. As in the case of the field square, in formulae (%
\ref{Tik}) and (\ref{T21}) we introduce a cutoff function and apply formula (%
\ref{sumformula}) for the summation over $n$. This enables us to present the
vacuum energy-momentum tensor in the form of the sum
\begin{equation}
\langle 0|T_{i}^{k}|0\rangle =\langle 0_{\mathrm{w}}|T_{i}^{k}|0_{\mathrm{w}%
}\rangle +\langle T_{i}^{k}\rangle _{\mathrm{cyl}},  \label{Tik1}
\end{equation}%
where $\langle 0_{\mathrm{w}}|T_{i}^{k}|0_{\mathrm{w}}\rangle $ is the part
corresponding to the geometry of a wedge without a cylindrical boundary and $%
\langle T_{i}^{k}\rangle _{\mathrm{cyl}}$ is induced by the cylindrical
shell. By taking into account (\ref{Hash1}), the latter may be written in
the form (no summation over $i$)
\begin{eqnarray}
\langle T_{i}^{i}\rangle _{\mathrm{cyl}} &=&\frac{q}{2\pi ^{2}}%
\sideset{}{'}{\sum}_{m=0}^{\infty }\sum_{\lambda =0,1}\int_{0}^{\infty
}dxx^{3}\frac{K_{qm}^{(\lambda )}(xa)}{I_{qm}^{(\lambda )}(xa)}F^{(i)}[\Phi
_{qm}^{(\lambda )}(\phi ),I_{qm}(xr)],  \label{Tikb} \\
\langle T_{2}^{1}\rangle _{\mathrm{cyl}} &=&\frac{q^{2}}{4\pi ^{2}}\frac{%
\partial }{\partial r}\sideset{}{'}{\sum}_{m=0}^{\infty }m\sin (2qm\phi
)\sum_{\lambda =0,1}(-1)^{\lambda }\int_{0}^{\infty }dxx\frac{%
K_{qm}^{(\lambda )}(xa)}{I_{qm}^{(\lambda )}(xa)}I_{qm}^{2}(xr),
\label{T21b}
\end{eqnarray}%
with the notations
\begin{eqnarray}
F^{(i)}[\Phi (\phi ),f(y)] &=&\Phi ^{2}(\phi )f^{2}(y),\;i=0,3,  \label{Fnu0}
\\
F^{(i)}[\Phi (\phi ),f(y)] &=&-(-1)^{i}\Phi ^{2}(\phi )f^{\prime 2}(y)-\left[
\Phi ^{2}(\phi )-(-1)^{i}\Phi ^{\prime 2}(\phi )/y^{2}\right]
f^{2}(y),\;i=1,2.  \label{Fnui}
\end{eqnarray}%
The diagonal components are symmetric with respect to the half-plane $\phi
=\phi _{0}/2$, whereas the off-diagonal component is an odd function under
the replacement $\phi \rightarrow \phi _{0}-\phi $. As it can be easily
checked, the tensor $\langle T_{i}^{k}\rangle _{\mathrm{cyl}}$ is traceless
and satisfies the covariant continuity equation $\nabla _{k}\langle
T_{i}^{k}\rangle _{\mathrm{cyl}}=0$. For the geometry under consideration
the latter leads to the relations%
\begin{eqnarray}
\frac{\partial }{\partial r}\left( r\langle T_{2}^{1}\rangle _{\mathrm{cyl}%
}\right) +r\frac{\partial }{\partial \phi }\langle T_{2}^{2}\rangle _{%
\mathrm{cyl}} &=&0,  \label{conteq1} \\
\frac{\partial }{\partial r}\left( r\langle T_{1}^{1}\rangle _{\mathrm{cyl}%
}\right) +r\frac{\partial }{\partial \phi }\langle T_{1}^{2}\rangle _{%
\mathrm{cyl}} &=&\langle T_{2}^{2}\rangle _{\mathrm{cyl}}.  \label{conteq2}
\end{eqnarray}%
As it is seen from formula (\ref{T21b}), the off-diagonal component $\langle
T_{2}^{1}\rangle _{\mathrm{cyl}}$ vanishes at the wedge sides and for these
points the VEV of the energy-momentum tensor is diagonal. By using the
inequalities $I_{\nu }^{\prime }(x)<\sqrt{1+\nu ^{2}/x^{2}}I_{\nu }(x)$ and $%
-K_{\nu }^{\prime }(x)>\sqrt{1+\nu ^{2}/x^{2}}K_{\nu }(x)$ for the modified
Bessel functions, it can be seen that $K_{\nu }^{\prime }(x)/I_{\nu
}^{\prime }(x)+K_{\nu }(x)/I_{\nu }(x)<0$. From this relation it follows
that the vacuum energy density induced by the cylindrical shell in the
interior region is always negative.

The renormalized VEV of the energy-momentum tensor for the geometry without
the cylindrical shell is obtained by using the corresponding formulae for
the field square. For the corresponding energy density one finds
\begin{equation}
\langle T_{0}^{0}\rangle _{\mathrm{w},\mathrm{ren}}=\frac{1}{8\pi }\left(
\langle E^{2}\rangle _{\mathrm{w},\mathrm{ren}}+\langle B^{2}\rangle _{%
\mathrm{w},\mathrm{ren}}\right) =-\frac{(q^{2}-1)(q^{2}+11)}{720\pi ^{2}r^{4}%
}.  \label{T00s}
\end{equation}%
As we see the parts in the VEVs of the field square which diverge on the
wedge sides cancel out and the corresponding energy density is finite
everywhere except the edge. Formula (\ref{T00s}) coincides with the
corresponding result for the geometry of the cosmic string (see \cite%
{Frol87,Dowk87}) with the angle deficit $2\pi -\phi _{0}$ and in the
corresponding formula $q=2\pi /\phi _{0}$. Other components are found from
the tracelessness condition and the continuity equation and one has \cite%
{jphy,Deutsch} (see also \cite{Most97})%
\begin{equation}
\langle T_{i}^{k}\rangle _{\mathrm{w},\mathrm{ren}}=-\frac{%
(q^{2}-1)(q^{2}+11)}{720\pi ^{2}r^{4}}\mathrm{diag}(1,1,-3,1).  \label{Tikw}
\end{equation}%
As we could expect this VEV vanishes for the geometry of a single plate
corresponding to $q=1$. In the limit $r\rightarrow \infty $, for fixed
values $x_{0}=r\phi _{0}$, from (\ref{Tikw}) the standard result for the
geometry of two parallel conducting plates is obtained.

The force acting on the wedge sides is determined by the component $\langle
T_{2}^{2}\rangle _{\mathrm{ren}}$ of the vacuum energy-momentum tensor
evaluated for $\phi =0$ and $\phi =\phi _{0}$. On the base of formula (\ref%
{Tik1}) for the corresponding effective pressure one has%
\begin{equation}
p_{2}=-\langle T_{2}^{2}\rangle _{\mathrm{ren}}|_{\phi =0,\phi _{0}}=p_{2%
\mathrm{w}}+p_{2\mathrm{cyl}},  \label{p2}
\end{equation}%
where%
\begin{equation}
p_{2\mathrm{w}}=-\frac{(q^{2}-1)(q^{2}+11)}{240\pi ^{2}r^{4}},  \label{p2w}
\end{equation}%
is the normal force acting per unit surface of the wedge for the case
without a cylindrical boundary and the additional term%
\begin{equation}
p_{2\mathrm{cyl}}=-\langle T_{2}^{2}\rangle _{\mathrm{cyl}}|_{\phi =0,\phi
_{0}}=-\frac{q}{\pi ^{2}}\sideset{}{'}{\sum}_{m=0}^{\infty }\sum_{\lambda
=0,1}\int_{0}^{\infty }dxx^{3}\frac{K_{qm}^{(\lambda )}(xa)}{%
I_{qm}^{(\lambda )}(xa)}F_{qm}^{(\lambda )}[I_{qm}(xr)],  \label{p2cyl}
\end{equation}%
with the notation%
\begin{equation}
F_{\nu }^{(\lambda )}[f(y)]=\left\{
\begin{array}{cc}
\nu ^{2}f^{2}(y)/y^{2}, & \lambda =0 \\
-f^{^{\prime }2}(y)-f^{2}(y), & \lambda =1%
\end{array}%
\right. ,  \label{Fnulam}
\end{equation}%
is induced by the cylindrical shell. From formula (\ref{p2w}) we see that
the corresponding vacuum forces are attractive for $q>1$ and repulsive for $%
q<1$. In particular, the equilibrium position corresponding to the geometry
of a single plate ($q=1$) is unstable. As regards to the part induced by the
cylindrical shell, from (\ref{p2cyl}) it follows that $p_{2\mathrm{cyl}}<0$
and, hence, the corresponding forces are always attractive.

Now, let us discuss the behavior of the boundary-induced part in the VEV of
the energy-momentum tensor in the asymptotic region of the parameters. Near
the cylindrical shell the main contribution comes from large values of $m$.
Thus, using the uniform asymptotic expansions for the modified Bessel
functions for large values of the order, up to the leading order, for the
points $a-r\ll a|\sin \phi |,a|\sin (\phi _{0}-\phi )|$ we find%
\begin{equation}
\langle T_{0}^{0}\rangle _{\mathrm{cyl}}\approx -\frac{1}{2}\langle
T_{2}^{2}\rangle _{\mathrm{cyl}}\approx -\frac{(a-r)^{-3}}{60\pi ^{2}a}%
,\;\langle T_{1}^{1}\rangle _{\mathrm{cyl}}\approx \frac{(a-r)^{-2}}{60\pi
^{2}a^{2}}.  \label{TiknearCyl}
\end{equation}%
These leading terms are the same as those for a cylindrical shell when the
wedge is absent. For the points near the edges $(r=a,\phi =0,\phi _{0})$ the
leading terms in the corresponding asymptotic expansions are the same as for
the geometry of a wedge with the opening angle $\phi _{0}=\pi /2$. The
latter are given by (\ref{Tikw}) with $q=2$. Near the edge, $r\rightarrow 0$%
, for the components (no summation over $i$) $\langle T_{i}^{i}\rangle _{%
\mathrm{cyl}}$, $i=0,3$, the main contribution comes from the mode $m=0$ and
we find%
\begin{equation}
\langle T_{i}^{i}\rangle _{\mathrm{cyl}}\approx \frac{q}{4\pi ^{2}a^{4}}%
\int_{0}^{\infty }dx\,x^{3}\frac{K_{0}^{\prime }(x)}{I_{0}^{\prime }(x)}%
=-0.0590\frac{q}{a^{4}},\,i=0,3.  \label{TiiNearEdgem0}
\end{equation}%
For the components (no summation over $i$) $\langle T_{i}^{i}\rangle _{%
\mathrm{cyl}}$, $i=1,2$, when $q>1$ the main contribution again comes form $%
m=0$ term and one has $\langle T_{i}^{i}\rangle _{\mathrm{cyl}}\approx
-\langle T_{0}^{0}\rangle _{\mathrm{cyl}}$, $i=1,2$. For $q<1$ the main
contribution into the components $\langle T_{i}^{i}\rangle _{\mathrm{cyl}}$,
$i=1,2$, comes from the term $m=1$ and we have (no summation over $i$)%
\begin{equation}
\langle T_{i}^{i}\rangle _{\mathrm{cyl}}\approx \frac{(-1)^{i}q\cos (2q\phi )%
}{2^{2q+1}\pi ^{2}\Gamma ^{2}(q)a^{4}}\left( \frac{r}{a}\right)
^{2(q-1)}\int_{0}^{\infty }dx\,x^{2q+1}\left[ \frac{K_{q}(x)}{I_{q}(x)}-%
\frac{K_{q}^{\prime }(x)}{I_{q}^{\prime }(x)}\right] ,\;i=1,2.
\label{TiiNearEdge}
\end{equation}%
In this case the radial and azimuthal stresses induced by the cylindrical
shell diverge on the edge $r=0$. In the case $q=1$ the sum of the
contributions of the terms with $m=0$ and $m=1$ given by formulae (\ref%
{TiiNearEdgem0}) and (\ref{TiiNearEdge}) should be taken. For the
off-diagonal component the main contribution comes from the $m=1$ mode with
the leading term%
\begin{equation}
\langle T_{2}^{1}\rangle _{\mathrm{cyl}}\approx \frac{q\sin (2q\phi )}{%
2^{2q+1}\pi ^{2}\Gamma ^{2}(q)a^{3}}\left( \frac{r}{a}\right)
^{2q-1}\int_{0}^{\infty }dx\,x^{2q+1}\left[ \frac{K_{q}(x)}{I_{q}(x)}-\frac{%
K_{q}^{\prime }(x)}{I_{q}^{\prime }(x)}\right] ,  \label{T21NearEdge}
\end{equation}%
and this component vanishes on the edge for $q>1/2$.

In the limit $q\gg 1$, the contribution of the modes with $m\geqslant 1$ is
suppressed by the factor $\exp [-2qm\ln (a/r)]$ and the main contribution
comes from the $m=0$ mode. The leading terms are given by the formulae (no
summation over $i$)%
\begin{eqnarray}
\langle T_{i}^{i}\rangle _{\mathrm{cyl}} &\approx &\frac{q}{4\pi ^{2}a^{4}}%
\int_{0}^{\infty }dx\,x^{3}\frac{K_{0}^{\prime }(x)}{I_{0}^{\prime }(x)}%
I_{0}^{2}(xr/a),\;i=0,3,  \label{Tii0qlarge} \\
\langle T_{i}^{i}\rangle _{\mathrm{cyl}} &\approx &-\frac{q}{4\pi ^{2}a^{4}}%
\int_{0}^{\infty }dx\,x^{3}\frac{K_{0}^{\prime }(x)}{I_{0}^{\prime }(x)}%
\left[ I_{0}^{2}(xr/a)+(-1)^{i}I_{1}^{2}(xr/a)\right] ,\;i=1,2.
\label{Tii1qlarge}
\end{eqnarray}%
Though in this limit the vacuum densities are large, due to the factor $1/q$
in the spatial volume, the corresponding global quantities tend to finite
value. In particular, as it follows from Eq. (\ref{Tii1qlarge}), in the
limit under consideration one has $\langle T_{i}^{i}\rangle _{\mathrm{cyl}%
}>0 $. Note that in the same limit the parts corresponding to the wedge
without the cylindrical shell behave as $q^{4}$ and, hence, for points not
too close to the shell these parts dominate in the VEVs.

\begin{figure}[tbph]
\begin{center}
\epsfig{figure=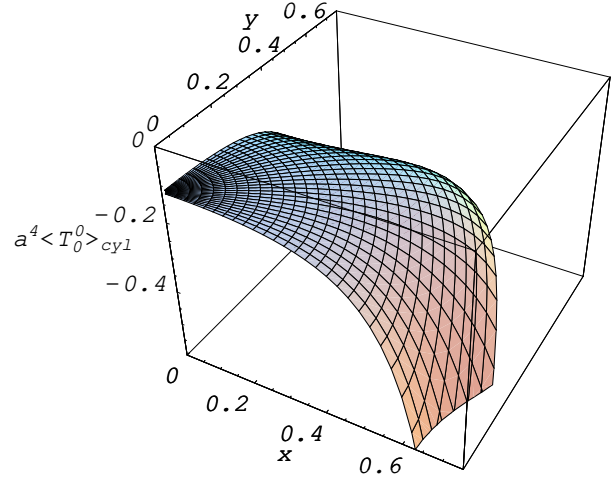,width=8.cm,height=6.5cm}
\end{center}
\caption{The part in the VEV of the energy density, $a^{4}\langle
T_{0}^{0}\rangle _{\mathrm{cyl}}$, induced by the cylindrical boundary as a
function on $x=(r/a)\cos \protect\phi $ and $y=(r/a)\sin \protect\phi $ for
a wedge with $\protect\phi _{0}=\protect\pi /2$. }
\label{fig2}
\end{figure}

\begin{figure}[tbph]
\begin{center}
\epsfig{figure=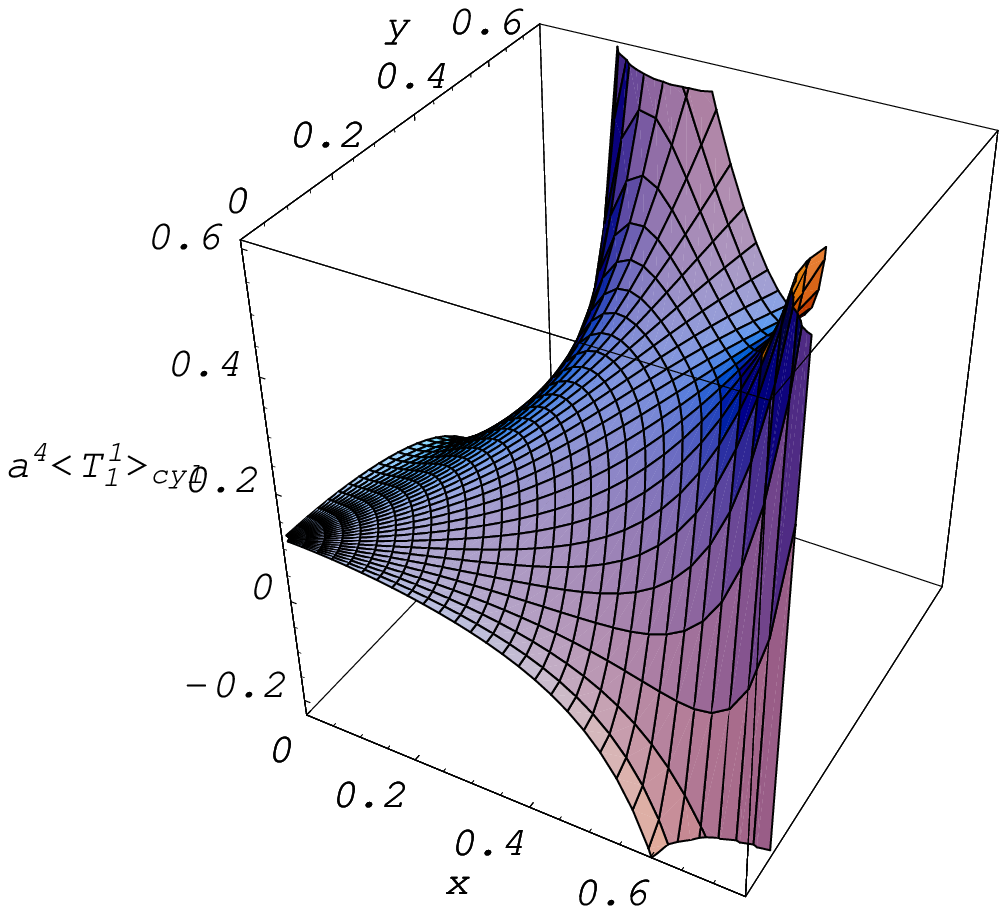,width=8.cm,height=6.5cm}
\end{center}
\caption{The part in the VEV of the radial stress, $a^{4}\langle
T_{1}^{1}\rangle _{\mathrm{cyl}}$, induced by the cylindrical boundary as a
function on $x=(r/a)\cos \protect\phi $ and $y=(r/a)\sin \protect\phi $ for
a wedge with $\protect\phi _{0}=\protect\pi /2$. }
\label{fig3}
\end{figure}

\begin{figure}[tbph]
\begin{center}
\epsfig{figure=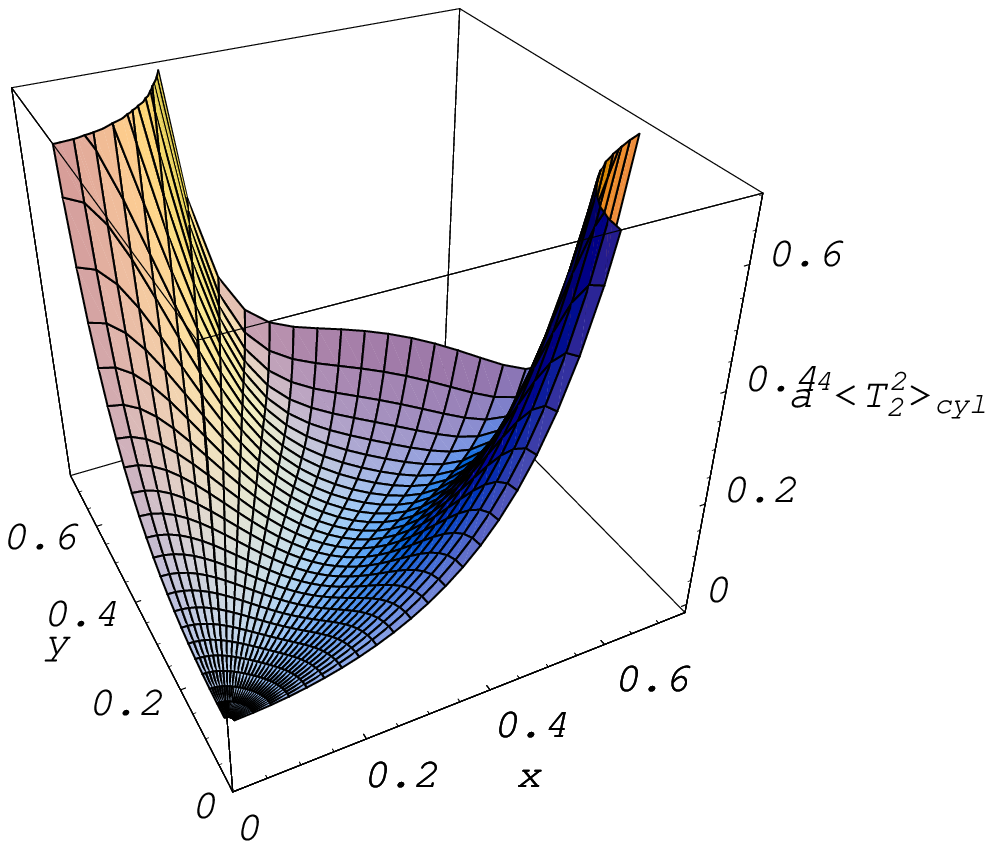,width=8.cm,height=6.5cm}
\end{center}
\caption{The part in the VEV of the azimuthal stress, $a^{4}\langle
T_{2}^{2}\rangle _{\mathrm{cyl}}$, induced by the cylindrical boundary as a
function on $x=(r/a)\cos \protect\phi $ and $y=(r/a)\sin \protect\phi $ for
a wedge with $\protect\phi _{0}=\protect\pi /2$. }
\label{fig4}
\end{figure}

\begin{figure}[tbph]
\begin{center}
\epsfig{figure=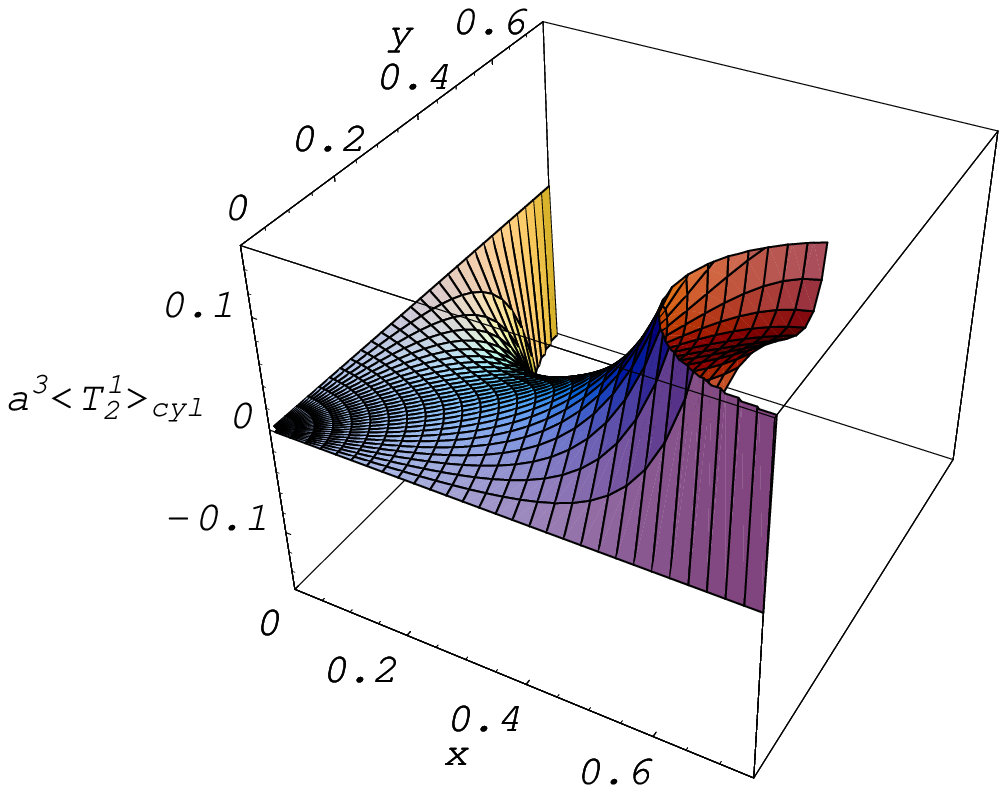,width=8.cm,height=6.5cm}
\end{center}
\caption{The part in the VEV of the off-diagonal component, $a^{4}\langle
T_{1}^{2}\rangle _{\mathrm{cyl}}$, induced by the cylindrical boundary as a
function on $x=(r/a)\cos \protect\phi $ and $y=(r/a)\sin \protect\phi $ for
a wedge with $\protect\phi _{0}=\protect\pi /2$. }
\label{fig5}
\end{figure}

In figures \ref{fig2}-\ref{fig5} we have plotted the parts in the VEVs of
the energy-momentum tensor induced by the cylindrical shell, $a^{4}\langle
T_{i}^{k}\rangle _{\mathrm{cyl}}$, as functions of $x=(r/a)\cos \phi $ and $%
y=(r/a)\sin \phi $, for a wedge with the opening angle $\phi _{0}=\pi /2$.

\begin{figure}[tbph]
\begin{center}
\epsfig{figure=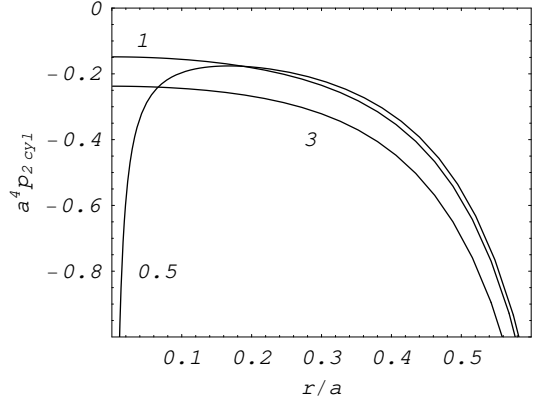,width=7.cm,height=5.5cm}
\end{center}
\caption{The effective azimuthal pressure induced by the cylindrical shell
on the wedge sides, $a^4p_{2\mathrm{cyl}}$, as a function of $r/a$. The
numbers near the curves correspond to the values of the parameter $q$.}
\label{fig6}
\end{figure}

In figure \ref{fig6} we have presented the dependence of the effective
azimuthal pressure induced by the cylindrical shell on the wedge sides, $%
a^4p_{2\mathrm{cyl}}$, as a function of $r/a$ for different values of the
parameter $q$.

There are several special cases of interest for the geometry of boundaries
we have considered. The case $\phi _{0}=\pi $ corresponds to the
semi-circular cylinder. The Casimir energy for the corresponding interior
region is evaluated in Ref. \cite{Nest01} by using the zeta function
technique. The case $\phi _{0}=2\pi $ corresponds to the geometry of a
cylindrical shell with a coaxial half-plane. And finally, the limit $\phi
_{0}\rightarrow 0$, $r,a\rightarrow \infty $, assuming that $a-r$ and $a\phi
_{0}\equiv b$ are fixed, corresponds to the geometry of two parallel plates
separated by a distance $b$, perpendicularly intersected by the third plate.
In the latter case it is convenient to introduce rectangular coordinates $%
(x^{\prime 1},x^{\prime 2},x^{\prime 3})=(x,y,z)$ with the relations $x=a-r$%
, $y=r\phi $. The components of the tensors in these coordinates we will
denote by primes. The corresponding vacuum energy-momentum tensor is
presented in the form
\begin{equation}
\langle 0|T_{k}^{\prime i}|0\rangle =\langle T_{k}^{\prime i}\rangle
^{(0)}+\langle T_{k}^{\prime i}\rangle ^{(1)},  \label{Tiklim}
\end{equation}%
where $\langle T_{k}^{\prime i}\rangle ^{(0)}$ is the vacuum expectation
value in the region between two parallel plates located at $y=0$ and $y=a$
and $\langle T_{k}^{\prime i}\rangle ^{(1)}$ is induced by the intersecting
plate at $x=0$. The latter is related to the quantities investigated above
by formulae%
\begin{equation}
\langle T_{i}^{\prime i}\rangle ^{(1)}=\lim \,\langle T_{i}^{i}\rangle _{%
\mathrm{cyl}},\quad \langle T_{2}^{\prime 1}\rangle ^{(1)}=-\lim \frac{1}{a}%
\langle T_{2}^{1}\rangle _{\mathrm{cyl}},  \label{limtrans}
\end{equation}%
with $\lim $ corresponding to the limit $a\rightarrow \infty $, $\phi
_{0}\rightarrow 0$ for fixed $a-r$ and $a\phi _{0}$. Taking this limit in
the term with $m=0$ of formula (\ref{Tikb}) we replace the modified Bessel
functions by the leading terms of the corresponding asymptotic formulae for
large values of the argument and the integral is taken elementary. For the
terms with $m\neq 0$ in formulae (\ref{Tikb}), (\ref{T21b}) we note that in
the limit under consideration one has $q=\pi /\phi _{0}\rightarrow \infty $,
and the order of the Bessel modified functions tends to infinity.
Introducing a new integration variable $x\rightarrow qmx$, we can replace
these functions by their uniform asymptotic expansions for large values of
the order. After these replacements the integration and the further
summation over $m$ are done by using the formulae from \cite{Prud86a}.

\section{Vacuum densities in the exterior region}

\label{sec:exter}

In this section we consider the VEVs for the field square and the
energy-momentum tensor in the region outside the cylindrical boundary
(region II in figure \ref{fig1}). The corresponding eigenfunctions for the
vector potential are obtained from formulae (\ref{Aalpha}) by the replacement%
\begin{equation}
J_{qm}(\gamma r)\rightarrow g_{qm}^{(\lambda )}(\gamma a,\gamma
r)=J_{qm}(\gamma r)Y_{qm}^{(\lambda )}(\gamma a)-Y_{qm}(\gamma
r)J_{qm}^{(\lambda )}(\gamma a),  \label{extreplace}
\end{equation}%
where, as before, $\lambda =0,1$ correspond to the waves of the electric and
magnetic types, respectively. Now, the eigenvalues for $\gamma $ are
continuous and in the normalization condition (\ref{Anorm}) the
corresponding part on the right is presented by the delta function. As the
normalization integral diverges for $\gamma ^{\prime }=\gamma $, the main
contribution into the integral comes from large values of $r$ and we can
replace the cylindrical functions with the argument $\gamma r$ by their
asymptotic expressions for large values of the argument. By this way it can
be seen that the normalization coefficient in the exterior region is
determined by the relation%
\begin{equation}
\beta _{\alpha }^{-2}=\frac{8\pi }{q}\delta _{m}\gamma \omega \left[
J_{qm}^{(\lambda )2}(\gamma a)+Y_{qm}^{(\lambda )2}(\gamma a)\right] .
\label{betalfext}
\end{equation}%
Substituting the eigenfunctions into the corresponding mode-sum formula, for
the VEV of the field square one finds%
\begin{equation}
\langle 0|F^{2}|0\rangle =\frac{2q}{\pi }\sideset{}{'}{\sum}_{m=0}^{\infty
}\int_{-\infty }^{+\infty }dk\int_{0}^{\infty }d\gamma \sum_{\lambda =0,1}%
\frac{\gamma ^{3}}{\sqrt{k^{2}+\gamma ^{2}}}\frac{g^{(\eta _{F\lambda
})}[\Phi _{qm}^{(\lambda )}(\phi ),g_{qm}^{(\lambda )}(\gamma a,\gamma r)]}{%
J_{qm}^{(\lambda )2}(\gamma a)+Y_{qm}^{(\lambda )2}(\gamma a)},
\label{F2ext}
\end{equation}%
where the functions $g^{(\eta _{F\lambda })}\left[ \Phi _{qm}^{(\lambda
)}(\phi ),g_{qm}^{(\lambda )}(\gamma a,\gamma r)\right] $ are defined by
relations (\ref{gnulam1}), (\ref{gnulam2}) with the function $%
f(x)=g_{qm}^{(\lambda )}(\gamma a,x)$. To extract from this VEV the part
induced by the cylindrical shell, we subtract from the right-hand side the
corresponding expression for the wedge without the cylindrical boundary. The
latter is given by formula (\ref{F2s}). The corresponding difference can be
further evaluated by using the identity%
\begin{eqnarray}
\frac{g^{(\eta _{F\lambda })}[\Phi _{qm}^{(\lambda )}(\phi
),g_{qm}^{(\lambda )}(\gamma a,\gamma r)]}{J_{qm}^{(\lambda )2}(\gamma
a)+Y_{qm}^{(\lambda )2}(\gamma a)} &=&g^{(\eta _{F\lambda })}[\Phi
_{qm}^{(\lambda )}(\phi ),J_{qm}(\gamma r)]  \notag \\
&&-\frac{1}{2}\sum_{s=1}^{2}\frac{J_{qm}^{(\lambda )}(\gamma a)}{%
H_{qm}^{(s)(\lambda )}(\gamma a)}g^{(\eta _{F\lambda })}[\Phi
_{qm}^{(\lambda )}(\phi ),H_{qm}^{(s)}(\gamma r)],  \label{ident}
\end{eqnarray}%
where $H_{qm}^{(1,2)}(z)$ are the Hankel functions. In order to transform
the integral over $\gamma $ with the last term on the right of (\ref{ident}%
), in the complex plane $\gamma $ we rotate the integration contour by the
angle $\pi /2$ for the term with $s=1$ and by the angle $-\pi /2$ for the
term with $s=2$. Due to the well-known properties of the Hankel functions
the integrals over the corresponding parts of the circles of large radius in
the upper and lower half-planes vanish. After introducing the modified
Bessel functions and integrating over $k$ with the help of formula (\ref%
{Hash1}), we can write the VEVs of the field square in the form (\ref{F21}),
where the part induced by the cylindrical shell is given by the formula
\begin{equation}
\langle F^{2}\rangle _{\mathrm{cyl}}=\frac{2q}{\pi }\sideset{}{'}{\sum}%
_{m=0}^{\infty }\sum_{\lambda =0,1}\int_{0}^{\infty }dx\,x^{3}\frac{%
I_{qm}^{(\lambda )}(xa)}{K_{qm}^{(\lambda )}(xa)}G^{(\eta _{F\lambda
})}[\Phi _{qm}^{(\lambda )}(\phi ),K_{qm}(xr)].  \label{F2bext}
\end{equation}%
In this formula the functions $G^{(\eta _{F\lambda })}\left[ \Phi (\phi
),f(x)\right] $ are defined by formulae (\ref{Gnujtilde0}), (\ref{Gnujtilde1}%
). Comparing this result with formula (\ref{F2cyl}), we see that the
expressions for the shell-induced parts in the interior and exterior regions
are related by the interchange $I_{qm}\rightleftarrows K_{qm}$. The VEV (\ref%
{F2bext}) diverges on the cylindrical shell with the leading term being the
same as that for the interior region. At large distances from the
cylindrical shell we introduce a new integration variable $y=xr$ and expand
the integrand over $a/r$. For $q>1$ the main contribution comes from the
lowest mode $m=0$ and up to the leading order we have
\begin{equation}
\langle E^{2}\rangle _{\mathrm{cyl}}\approx \frac{4q}{5\pi r^{4}}\left(
\frac{a}{r}\right) ^{2},\;\langle B^{2}\rangle _{\mathrm{cyl}}\approx -\frac{%
28q}{15\pi r^{4}}\left( \frac{a}{r}\right) ^{2}.  \label{B2far}
\end{equation}%
For $q<1$ the dominant contribution into the VEVs at large distances is due
to the mode $m=1$ with the leading term%
\begin{equation}
\langle F^{2}\rangle _{\mathrm{cyl}}\approx -\frac{4q^{2}(q+1)}{\pi r^{4}}%
\left( \frac{a}{r}\right) ^{2q}\left[ \frac{\cos (2q\phi )}{2q+3}+(-1)^{\eta
_{F1}}\frac{q+1}{2q+1}\right] .  \label{F2far}
\end{equation}%
For the case $q=1$ the contributions of the modes $m=0$ and $m=1$ are of the
same order and the corresponding leading terms are obtained by summing these
contributions. The latter are given by the right-hand sides of formulae (\ref%
{B2far}) and (\ref{F2far}). As we see, at large distances the part induced
by the cylindrical shell is suppressed with respect to the part
corresponding to the wedge without the shell by the factor $(a/r)^{2\beta }$
with $\beta =\min (1,q)$.

Now we turn to the VEVs of the energy-momentum tensor in the exterior
region. Substituting the eigenfunctions into the corresponding mode-sum
formula, one finds (no summation over $i$)%
\begin{eqnarray}
\langle 0|T_{i}^{i}|0\rangle &=&\frac{q}{4\pi ^{2}}\sideset{}{'}{\sum}%
_{m=0}^{\infty }\int_{-\infty }^{+\infty }dk\int_{0}^{\infty }d\gamma
\sum_{\lambda =0,1}\frac{\gamma ^{3}}{\sqrt{k^{2}+\gamma ^{2}}}\frac{%
f^{(i)}[\Phi _{qm}^{(\lambda )}(\phi ),g_{qm}^{(\lambda )}(\gamma a,\gamma
r)]}{J_{qm}^{(\lambda )2}(\gamma a)+Y_{qm}^{(\lambda )2}(\gamma a)},
\label{Tikext0} \\
\langle 0|T_{2}^{1}|0\rangle &=&-\frac{q}{8\pi ^{2}}\frac{\partial }{%
\partial r}\sideset{}{'}{\sum}_{m=0}^{\infty }m\sin (2qm\phi )\int_{-\infty
}^{+\infty }dk\int_{0}^{\infty }d\gamma \sum_{\lambda =0,1}(-1)^{\lambda }%
\frac{\gamma g_{qm}^{(\lambda )2}(\gamma a,\gamma r)}{\sqrt{k^{2}+\gamma ^{2}%
}}.  \label{T21ext0}
\end{eqnarray}%
Subtracting from these VEVs the corresponding expression for the wedge
without the cylindrical boundary, analogously to the case of the field
square, it can be seen that the VEVs are presented in the form (\ref{Tik1}),
with the parts induced by the cylindrical shell given by the formulae (no
summation over $i$)
\begin{eqnarray}
\langle T_{i}^{i}\rangle _{\mathrm{cyl}} &=&\frac{q}{2\pi ^{2}}%
\sideset{}{'}{\sum}_{m=0}^{\infty }\sum_{\lambda =0,1}\int_{0}^{\infty
}dxx^{3}\frac{I_{qm}^{(\lambda )}(xa)}{K_{qm}^{(\lambda )}(xa)}F^{(i)}[\Phi
_{qm}^{(\lambda )}(\phi ),K_{qm}(xr)],  \label{Tikbext} \\
\langle T_{2}^{1}\rangle _{\mathrm{cyl}} &=&\frac{q^{2}}{4\pi ^{2}}\frac{%
\partial }{\partial r}\sideset{}{'}{\sum}_{m=0}^{\infty }m\sin (2qm\phi
)\sum_{\lambda =0,1}(-1)^{\lambda }\int_{0}^{\infty }dxx\frac{%
I_{qm}^{(\lambda )}(xa)}{K_{qm}^{(\lambda )}(xa)}K_{qm}^{2}(xr).
\label{T21bext}
\end{eqnarray}%
Here the functions $F^{(i)}\left[ \Phi (\phi ),f(y)\right] $ are defined by
formulae (\ref{Fnu0}), (\ref{Fnui}). By using the inequality given in the
paragraph after formula (\ref{conteq2}), we can show that the vacuum energy
density induced by the cylindrical shell in the exterior region is positive.

In the way similar to that for the interior region, for the force acting on
the wedge sides is presented in the form of the sum (\ref{p2}), where the
part corresponding to the wedge without a cylindrical shell is determined by
formula (\ref{p2w}) and for the part due to the presence of the cylindrical
shell we have%
\begin{equation}
p_{2\mathrm{cyl}}=-\langle T_{2}^{2}\rangle _{\mathrm{cyl}}|_{\phi =0,\phi
_{0}}=-\frac{q}{\pi ^{2}}\sideset{}{'}{\sum}_{m=0}^{\infty }\sum_{\lambda
=0,1}\int_{0}^{\infty }dxx^{3}\frac{I_{qm}^{(\lambda )}(xa)}{%
K_{qm}^{(\lambda )}(xa)}F_{qm}^{(\lambda )}[K_{qm}(xr)].  \label{p2cylext}
\end{equation}%
In this formula, the function $F_{\nu }^{(\lambda )}\left[ f(y)\right] $ is
defined by relations (\ref{Fnulam}) and the corresponding forces are always
attractive.

\begin{figure}[tbph]
\begin{center}
\epsfig{figure=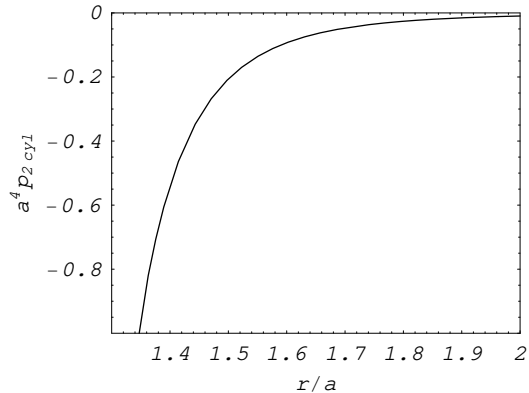,width=7.cm,height=5.5cm}
\end{center}
\caption{The effective azimuthal pressure induced by the cylindrical shell
on the wedge sides, $a^4p_{2\mathrm{cyl}}$, as a function of $r/a$ in the
exterior region for $q=1$. The curves for the values $q=0.5,3$ are close to
the plotted one.}
\label{fig7}
\end{figure}

The leading divergence in the boundary induced part (\ref{Tikbext}) on the
cylindrical surface is given by the same formulae as for the interior
region. For large distances from the shell and for $q>1$ the main
contribution into the VEVs of the diagonal components comes from the $m=0$, $%
\lambda =1$ term and one has (no summation over $i$)%
\begin{equation}
\langle T_{i}^{i}\rangle _{\mathrm{cyl}}\approx -\frac{qc_{i}}{15\pi
^{2}r^{4}}\left( \frac{a}{r}\right)
^{2},\;c_{0}=c_{3}=2,\;c_{1}=1,\;c_{2}=-5.  \label{Tikfar}
\end{equation}%
In the case $q<1$ the main contribution into the VEVs of the diagonal
components at large distances from the cylindrical shell comes from the $m=1$
mode. The leading terms in the corresponding asymptotic expansions are given
by the formulae%
\begin{equation}
\langle T_{i}^{i}\rangle _{\mathrm{cyl}}\approx -q^{2}(q+1)c_{i}(q)\frac{%
\cos (2q\phi )}{\pi ^{2}r^{4}}\left( \frac{a}{r}\right) ^{2q},
\label{Tiifar1}
\end{equation}%
with the notations%
\begin{equation}
c_{0}(q)=c_{3}(q)=\frac{1}{2q+3},\;c_{1}(q)=\frac{2q^{2}+q+1}{(2q+1)(2q+3)}%
,\;c_{2}(q)=-\frac{q+1}{2q+1}.  \label{ci(q)}
\end{equation}%
In the case $q=1$ the asymptotic terms are determined by the sum of the
contributions coming from the modes $m=0$ and $m=1$. The latter are given by
formulae (\ref{Tikfar}), (\ref{Tiifar1}). For the off-diagonal component,
for all values $q$ the main contribution at large distances comes from the $%
m=1$ mode with the leading term%
\begin{equation}
\langle T_{2}^{1}\rangle _{\mathrm{cyl}}\approx -\frac{q^{3}(q+1)}{2q+1}%
\frac{\sin (2q\phi )}{\pi ^{2}r^{3}}\left( \frac{a}{r}\right) ^{2q}.
\label{T21far}
\end{equation}%
For large values of $q$, $q\gg 1$, the contribution of the terms with $m>0$
is suppressed by the factor $\exp [-2qm\ln (r/a)]$ and the main contribution
comes form the $m=0$ term with the behavior $\langle F^{2}\rangle _{\mathrm{%
cyl}}\propto q$ and $\langle T_{i}^{k}\rangle _{\mathrm{cyl}}\propto q$. In
figure \ref{fig7} we have plotted the dependence of the effective azimuthal
pressure induced by the cylindrical shell on the wedge sides, $a^{4}p_{2%
\mathrm{cyl}}$, as a function of $r/a$ for $q=1$.

\section{Conclusion}

\label{sec:Conc}

In this paper we have investigated the polarization of the
electromagnetic vacuum by a wedge with a coaxial cylindrical
boundary, assuming that all boundaries are perfectly conducting.
Both regions inside and outside of the cylindrical shell (regions
I and II in figure \ref{fig1}) are considered. In section
\ref{sec:Inter} we have evaluated the VEVs of the field square in
the interior region. The corresponding mode-sums contain series
over the zeros of the Bessel function for TM modes and its
derivative for TE modes. For the summation of these series we used
a variant of the generalized Abel-Plana formula. The latter
enables us to extract from the VEVs the parts corresponding to the
geometry of a wedge without a cylindrical shell and to present the
parts induced by the shell in terms of integrals which are
exponentially convergent for points away from the boundaries. For
the wedge without the cylindrical shell the VEVs of the field
square are presented in the form (\ref{F2wren}). The first term on
the right of this formula corresponds to the VEVs for the geometry
of a cosmic string with the angle deficit $2\pi -\phi _{0}$. The
angle-dependent parts in the VEVs of the electric and magnetic
fields have opposite signs and are cancelled in the evaluation of
the vacuum energy density. The parts induced by the cylindrical
shell are presented in the form (\ref{F2cyl}). We have discussed
this general formula in various asymptotic regions of the
parameters
including the points near the edges and near the shell. In section \ref%
{sec:EMTint} we consider the VEV of the energy-momentum tensor in the region
inside the shell. As for the field square, the application of the Abel-Plana
formula allows us to present this VEV in the form of the sum of purely wedge
and shell-induced parts, formula (\ref{Tik1}). For the geometry of a wedge
without the cylindrical boundary the vacuum energy-momentum tensor\ does not
depend on the angle $\phi $ and is the same as in the geometry of the cosmic
string and is given by formula (\ref{Tikw}). The corresponding vacuum forces
acting on the wedge sides are attractive for $\phi _{0}<\pi $ and repulsive
for $\phi _{0}>\pi $. In particular, the equilibrium position corresponding
to the geometry of a single plate ($\phi _{0}=\pi $) is unstable. For the
region inside the shell the part in the VEV of the energy-momentum tensor
induced by the presence of the cylindrical shell is non-diagonal and the
corresponding components are given by formulae (\ref{Tikb}), (\ref{T21b}).
The vacuum energy density induced by the cylindrical shell in the interior
region is negative. We have investigated the vacuum densities induced by the
cylindrical shell in various asymptotic regions of the parameters. For
points near the cylindrical shell the leading terms in the asymptotic
expansions over the distance from the shell are given by formulae (\ref%
{TiknearCyl}). These terms are the same as those for a cylindrical shell
when the wedge is absent. For a wedge with $\phi _{0}<\pi $ the part in the
vacuum energy-momentum tensor induced by the shell is finite on the edge $%
r=0 $. For $\phi _{0}>\pi $ the shell-induced parts in the energy density
and the axial stress remain finite, whereas the radial and azimuthal
stresses diverge as $r^{2(\pi /\phi _{0}-1)}$. The corresponding
off-diagonal component behaves like $r^{2\pi /\phi _{0}-1}$ for all values $%
\phi _{0}$. For the points near the edges $(r=a,\phi =0,\phi _{0})$ the
leading terms in the corresponding asymptotic expansions are the same as for
the geometry of a wedge with the opening angle $\phi _{0}=\pi /2$. In the
limit of small opening angles, $\phi _{0}\ll \pi $, the shell-induced parts
behave like $1/\phi _{0}$. In the same limit the parts corresponding to the
wedge without the shell behave as $1/\phi _{0}^{4}$, and for points not too
close to the shell these parts dominate in the VEV of the energy-momentum
tensor. The presence of the shell leads to additional forces acting on the
wedge sides. The corresponding effective azimuthal pressure is given by
formula (\ref{p2cyl}) and these forces are always attractive.

The VEVs of the field square and the energy-momentum tensor in the region
outside the cylindrical shell are investigated in section \ref{sec:exter}.
As in the case of the interior region, these VEVs are presented as sums of
the parts corresponding to the wedge without the cylindrical shell and the
parts induced by the shell. The latter are given by formula (\ref{F2bext})
for the field square and by formulae (\ref{Tikbext}), (\ref{T21ext0}) for
the components of the energy-momentum tensor. In the exterior region the
vacuum energy density induced by the cylindrical shell is always positive.
Additional forces acting on the wedge sides due to the presence of the shell
are given by formula (\ref{p2cylext}). As in the case of the interior region
these forces are attractive. For large values of the parameter $q$, the
contribution into the parts induced by the cylindrical shell coming from the
modes with $m\neq 0$ is exponentially suppressed, whereas the contribution
of the lowest mode $m=0$ is proportional to $q$. Though in this limit the
vacuum densities are large, due to the factor $1/q$ in the spatial volume
element, the corresponding global quantities tend to finite limiting values.

\section*{Acknowledgments}

The work was supported by the Armenian Ministry of Education and Science
Grant No. 0124.

\end{document}